\documentclass[aoas,MSNbibl,nameyear,dvips]{arximspdf}
\usepackage{dcolumn}
\usepackage{graphicx}

\doi{10.1214/13-AOAS673} 
\volume{7}
\issue{4}
\pubyear{2013}
\firstpage{2315}
\lastpage{2335}

\makeatletter

\newcolumntype{d}[1]{D{.}{.}{#1}}

\renewcommand{\mid}{|}

\def\R{\mathbb{R}}

\def\aa{\alpha_1}

\def\la{\lambda}

\def\bY{\mathbf{Y}}
\def\bg{{\bolds{\gamma}}}

\newcommand{\argmax}{\mathop{\arg\max}}

\newtheorem{theorem}{Theorem}

\makeatother

\begin{document}
\begin{frontmatter}

\title{Fitting birth--death processes to panel data with applications to
bacterial DNA fingerprinting}
\runtitle{Birth--death processes for panel data}

\begin{aug}
\author[A]{\fnms{Charles R.} \snm{Doss}\thanksref{t7}\ead[label=e1]{cdoss@uw.edu}},
\author[B]{\fnms{Marc A.} \snm{Suchard}\thanksref{t2,t6,t8}\ead[label=e2]{msuchard@ucla.edu}},
\author[C]{\fnms{Ian} \snm{Holmes}\thanksref{t3,t9}\ead[label=e3]{ihh@berkeley.edu}},\\
\author[D]{\fnms{Midori} \snm{Kato-Maeda}\thanksref{t4,t10}\ead[label=e4]{Midori.Kato-Maeda@ucsf.edu}}
\and
\author[A]{\fnms{Vladimir N.} \snm{Minin}\corref{}\thanksref{t5,t6,t7}\ead[label=e5]{vminin@uw.edu}}
\runauthor{C. R. Doss et al.}
\affiliation{University of Washington, Seattle\thanksmark{t7},
University of California, Los Angeles\thanksmark{t8},
University of California, Berkeley\thanksmark{t9},
and University of California, San~Francisco\thanksmark{t10}}
\address[A]{C. R. Doss\\
V. N. Minin\\
Department of Statistics \\
University of Washington\\
Seattle, Washington 98195\\
USA\\
\printead{e1}\\
\hphantom{E-mail: }\printead*{e5}}
\address[B]{M. A. Suchard\\
Departments of Biomathematics,\\
\quad Biostatistics and Human Genetics\\
University of California, Los Angeles\hspace*{8.2pt}\\
Los Angeles, California 90095\\
USA\\
\printead{e2}}
\address[C]{I. Holmes\\
Department of Bioengineering \\
\quad and Biophysics Graduate Group\\
University of California, Berkeley\\
Berkeley, California 94720\\
USA\\
\printead{e3}}
\address[D]{M. Kato-Maeda\\
Department of Medicine\\
San Francisco General Hospital\\
University of California, San Francisco\\
San Francisco, California 94143\\
USA\\
\printead{e4}}
\end{aug}

\thankstext{t2}{Supported by NIH Grant R01 GM086887.}

\thankstext{t3}{Supported by NIH Grant GM076705.}

\thankstext{t4}{Supported by NIH Grant AI034238.}

\thankstext{t5}{Supported by the UW Royalty Research Fund.}

\thankstext{t6}{Supported by NSF Grant DMS-08-56099.}

\received{\smonth{12} \syear{2012}}
\revised{\smonth{7} \syear{2013}}

%
\begin{abstract}
Continuous-time linear birth--death-immigration (BDI) processes are
frequently used in ecology and epidemiology to model stochastic
dynamics of the population of interest. In clinical settings, multiple
birth--death processes can describe disease trajectories of individual
patients, allowing for estimation of the effects of individual
covariates on the birth and death rates of the process. Such estimation
is usually accomplished by analyzing patient data collected at unevenly
spaced time points, referred to as panel data in the biostatistics
literature. Fitting linear BDI processes to panel data is a nontrivial
optimization problem because birth and death rates can be functions of
many parameters related to the covariates of interest. We propose a
novel expectation--maximization (EM) algorithm for fitting linear BDI
models with covariates to panel data. We derive a closed-form
expression for the joint generating function of some of the BDI process
statistics and use this generating function to reduce the E-step of the
EM algorithm, as well as calculation of the Fisher information, to
one-dimensional integration. This analytical technique yields a
computationally efficient and robust optimization algorithm that we
implemented in an open-source \texttt{R} package. We apply our method to
DNA fingerprinting of \textit{Mycobacterium tuberculosis}, the
causative agent of tuberculosis, to study intrapatient time evolution
of IS\textit{6110} copy number, a~genetic marker frequently used during
estimation of epidemiological clusters of \textit{Mycobacterium
tuberculosis} infections. Our analysis reveals previously undocumented
differences in IS\textit{6110} birth--death rates among three major
lineages of \textit{Mycobacterium tuberculosis}, which has important
implications for epidemiologists that use IS\textit{6110} for DNA
fingerprinting of \textit{Mycobacterium tuberculosis}.
\end{abstract}

%

\end{frontmatter}

\section{Introduction}


Linear birth--death-immigration (BDI) processes provide useful building
blocks for modeling population dynamics in ecology [\citet{Nee2006BDlitReview}],
molecular evolution [\citet{TKF91}] and epidemiology
[\citet{GibsonRenshaw1998bd}], among many other areas.
Although \citet{Keiding1975BD} has extensively studied inference for
fully observed
continuous-time BDI processes, more often such processes are not
observed completely, posing
challenging computational problems for statisticians. Here, we use
applied probability tools to develop a new, efficient implementation of
the expectation--maximization (EM) algorithm for fitting discretely
observed BDI processes.

We are interested in situations where we observe multiple independent
conti\-nuous-time BDI trajectories at fixed, possibly irregularly spaced, time
points. Such observations, called panel data, often arise in medical
applications, with independent BDI trajectories corresponding to some
stochastic process recorded in different patients under study [\citet
{Crespi2005}]. The
birth and death rates can then be modeled as functions of patient-specific
covariates. This modeling framework is similar to the use of continuous-time
Markov chains (CTMCs) in multi-state disease progression models with a finite
number of states [\citet{Kalbfleisch1985}]. Although established
methods for
fitting finite state CTMCs to panel data exist [\citet{Kalbfleisch1985,Lange1995gradientEM,Jackson2011}], less attention has been paid to infinite
state-space processes, such as BDI models.

Outside of medical applications, estimating parameters of discretely observed
BDI models is considered in the molecular evolution and bioinformatics
literature [\citet{TKF91,Holmes2005EMindel}]. For example,
\citet{Holmes2005EMindel} proposed an EM algorithm for discretely
observed BDI
processes in the context of finding the most optimal alignment of multiple
genomic sequences.
The author argues that the EM algorithm's simplicity and robustness
make this
method attractive for large-scale bioinformatics applications. Unfortunately,
implementation of the EM algorithm by \citet{Holmes2005EMindel} is applicable
only to a very restricted class of BDI processes. In this paper, we
develop a
more general EM algorithm that applies to a large class of BDI models
and is
not restricted to molecular evolution applications.

Computing expectations of the complete-data log-likelihood, needed for
executing an EM algorithm, can be challenging, especially if the
complete-data were generated by a continuous-time stochastic process. When
the complete data are generated by a finite state-space CTMC, these
expectations can be computed efficiently
[\citet{Lange1995gradientEM,Holmes2002EMseqAlign}]. Although the BDI
process is
also a CTMC, the infinite state-space of the process prohibits us from using
these computationally efficient methods. \citet{Holmes2005EMindel} considers
a BDI model with the immigration rate either zero or proportional to
the birth
rate. Under this restriction, the complete-data likelihood belongs to the
exponential family, which means that the complete-data log-likelihood
is a
linear function of sufficient statistics of the complete data. Making
further stringent assumptions about the initial state of the process,
\citet{Holmes2005EMindel} computes expectations of these sufficient
statistics by numerically solving a system of coupled nonlinear ordinary
differential equations (ODEs). Working with this birth--death-restricted
immigration (BDRI)
model, but without any restrictions on the starting state of the
process, we
develop a new computationally efficient method for computing the expected
sufficient statistics. Our method combines ideas from \citet{Kendall1948}
and \citet{Lange1982fft} and reduces computations of the expected sufficient
statistics to one-dimensional integration, a~computational task that is much
simpler than solving a system of nonlinear ODEs. We develop a similar
integration method to compute the observed Fisher information matrix
via Louis' formula [\citet{Louis1982}]
and use this matrix for calculation of confidence intervals and sets. In
addition, when we have multiple BDRI trajectories observed, we allow
the birth
and death rates to be functions of trajectory-specific covariates.

We first test our EM algorithm on simulated data and then turn to a problem
of estimating birth and death rates of the transposable element
IS\textit{6110} in \textit{Mycobacterium tuberculosis}, the causative
bacterial agent of most tuberculosis (TB) in humans.
\textit{Mycobacterium tuberculosis} genome carries multiple IS\textit
{6110} copies
that get duplicated and deleted rapidly during replication.
Estimating IS\textit{6110} copy number birth (duplication) and death
(loss) rates is an important task in TB molecular epidemiology
because researchers use IS\textit{6110} copy number to group infected
individuals into epidemiological clusters [\citet{Small1994}]. In the United
States, the resurgence of TB cases, attributed to significant changes in
socioeconomic factors, started in the late 1980s, with the number of TB
cases reaching its peak in 1991 and steadily declining since then
[\citet{CattamanchiEtAl2006Tuberculosis}]. Since 1991, the University of
California, San Francisco has been maintaining a database of TB cases
reported to the San Francisco Department of Public Health. The database
contains demographic and certain clinical information as well as
\textit{M. tuberculosis} genotypes (e.g., IS\textit{6110}
copy number) for each reported TB case [\citet{Jasmer1999}].
\citet{RosenbergTrans03} used a subset of this database to estimate
IS\textit{6110} birth and death rates. These authors proposed an approximate
likelihood method to accomplish this estimation. We revisit this problem
using our EM algorithm and compare our results with the approximation of
\citet{RosenbergTrans03}. Further, we examine differences in birth and death
rates among three main lineages of \textit{M. tuberculosis} and find
that the
East-Asian \textit{M. tuberculosis} is evolving at a slower rate than its
European--American counterpart. This novel finding has serious implications
on the definition of epidemiological clusters based on the
IS\textit{6110} copy number. To investigate the
possibility of spurious effect of \textit{M. tuberculosis} lineage on
IS\textit{6110} birth and death rates due to a confounding factor, we
build a
more complicated model for birth and death rates. In addition to the lineage,
we include \textit{M. tuberculosis} drug-resistance status and HIV infection
status of each patient as birth and death rate covariates. We find that after
including these covariates, the lineage remains the only variable that
significantly affects IS\textit{6110} birth and death rates.

\section{BDRI process with covariates}
\label{secMethodology}
We start with $m$ independent conti\-nuous-time homogeneous linear BDRI processes
$\{ X_{p,t} \}$, for $p=1,\ldots, m$, with corresponding per capita
birth rates $\la_p \ge0$, per capita death rates
$\mu_p \ge0$ and immigration rates $\nu_p = \beta\la_p$, where
$\beta\ge
0$ is a known constant. Assuming that each process $p$ has $c_1$ covariates
related to the birth rates and $c_2$ covariates related to the death rates,
collected into vectors $\mathbf{z}_{p,\lambda}' = (z_{p,\lambda,1},\ldots,
z_{p,\lambda, c_1}) \in\mathbb{R}^{c_1}$ and $\mathbf{z}_{p,\mu}' =
(z_{p,\mu,1},\ldots, z_{p,\mu, c_2}) \in\mathbb{R}^{c_2}$, we
model birth
and death rates as log-linear functions of these covariates:
%
\begin{equation}
\label{eqparametrization} \log\lambda_p = \mathbf{z}_{p,\la}'
\bg_{\la} \quad\mbox{and}\quad \log\mu_p = \mathbf{z}_{p,\mu}'
\bg_{\mu},
\end{equation}
where $\bg_{\la}' = (\gamma_{\lambda,1},\ldots, \gamma_{\lambda,c_1})$ and
$\bg_{\mu}' = (\gamma_{\mu,1},\ldots, \gamma_{\mu,c_2})$ are
birth and death regression coefficients.
Covariate vectors $\mathbf{z}_{p,\lambda}$ and $\mathbf{z}_{p,\mu}$
are assumed to be known and fixed for every process $p$.
For example, if each BDRI process models a disease related trajectory
for each patient, then covariates are usually composed of
patient-specific clinical and demographic information (e.g., gender,
medical history).

We assume that we observe the $p$th process at $n(p)+1$ distinct times,
$0=t_{p,0} < t_{p,1} < \cdots< t_{p,n(p)}$. We denote our data vector by
\[
\bY= (X_{1,t_{1,0}},\ldots,X_{1,t_{1,n(1)}},\ldots, X_{m,t_{m,0}},\ldots,X_{m,t_{m,n(m)}} )
\]
and the parameter vector by $\bg= ( \bg_\la, \bg_\mu)
\in\mathbb{R}^{c_1+c_2}$.
We are interested in computing the parameter maximum likelihood
estimates (MLEs),
$\hat{\bg} =\break  \arg\max_{\bg}
l_o(\bY; \bg)$, where
%
\begin{equation}
\label{eqpartialLikelihood} l_o (\bY; \bg):= \sum
_{p=1}^m \sum_{i=0}^{n(p)-1}
\log p_{X_{p,t_{p,i}},X_{p,t_{p,i+1}}} ( t_{p,i+1}-t_{p,i}; \lambda_p,
\mu_p )
\end{equation}
is the observed-data log-likelihood and $p_{i,j} (t; \lambda,\mu
)= P_{\lambda,\mu} (X_t=j \mid X_0=i )$,
$i,j=0,1,\ldots\,$, are the transition probabilities of the BDRI process.
These transition probabilities can be calculated either using the
generating function derived by \citet{Kendall1948} or via the
orthogonal polynomial representation of
\citet{KarlinMcGregor1958summary}.
Despite the explicit algebraic nature of the orthogonal polynomials,
the latter method
can be numerically unstable and the generating function method is often
preferred [\citet{Sehl2009}].
Although one can maximize the likelihood $l_o (\bY; \bg
)$ using standard
off-the-shelf optimization algorithms, such generic algorithms can be
problematic
when the BDI rates are functions of a high-dimensional parameter
vector, such
as the vector of regression coefficients $\bg$ in our case.
As an alternative to generic optimization, we develop an EM algorithm, known
for its robustness and ability to cope with high-dimensional optimization
[\citet{DempsterLairdRubin1977EM}].

\section{EM algorithm for the BDRI process}
The complete data in our case consist of the BDRI trajectories $\{
X_{p,t}\}$,
observed continuously during the corresponding intervals $[0,t_{p,n(p)}]$,
$p=1,\ldots,m$. Let $\mathbf{X} = \{X_{p,t}\}_{p=1,\ldots,m}^{t \in
[0,t_{p, n(p)}]}$ be the
complete data and let $l_c(\mathbf{X}; \bg)$ be the complete data
log-likelihood.
The EM algorithm starts by initializing the parameter vector to an
arbitrarily chosen vector $\bg_0$.
At the $k$th iteration of the algorithm we set
%
\begin{equation}\label{emiter}
\bg_k = \argmax_{\bg} \mathrm{E}_{\bg_{k-1}} \bigl[
l_c(\mathbf{X}; \bg) \mid \bY\bigr].
\end{equation}
To accomplish the above maximization, we need to be able to evaluate
the expectation
in (\ref{emiter}) for any vector $\bg$. Traditionally, a numerical
procedure for computing such an expectation is
called an E-step of the EM algorithm. The maximization of the
expectation is called an M-step of
the EM algorithm.
Below, we develop efficient algorithms for implementing these E- and
M-steps for the discretely observed BDRI process.
As is often the case, we will see that to compute the needed
expectations for all $\bg\in\R^{c_1+c_2}$, we need to compute only the
expectations of certain statistics that do not depend on~$\bg$.

\subsection{E-step}
\label{secEM2param}
Since our BDRI process is a CTMC, the log-likelihood of the complete
data is
%
\begin{eqnarray}
l_c(\mathbf{X}; \bg) &=& - \sum_{p=1}^m
\Biggl[\sum_{i=0}^\infty d^p(i)
\bigl[i(\lambda_p + \mu_p) + \nu_p \bigr]
\nonumber\\[-8pt]\\[-8pt]
&&\hspace*{31.2pt}{} + \sum_{i=0}^\infty\bigl(n^p_{i,i+1}
\log(i \lambda_p + \nu_p) + n^p_{i,i-1}
\log(i \mu_p) \bigr) \Biggr],
\nonumber
\end{eqnarray}
where $d^p(i)$ is the total time spent by $X_{p,t}$ in state $i$ and
$n^p_{i,j}$ is the number of jumps from state $i$ to state $j$ during the
interval $[0,t_{p,n(p)}]$ [\citet{Guttorp1995book}]. Replacing $\nu_p$ with
$\beta\lambda_p$ in the above equation, we arrive at\vadjust{\goodbreak} a more compact
representation of the complete-data log-likelihood:
%
\begin{eqnarray}
\label{eq2paramFullLike} l_c(\mathbf{X}; \bg) &=& \sum
_{p=1}^m  \bigl[ - R_{p, t_{p,n(p)}} (
\lambda_p + \mu_p) - t_{n(p)} \beta
\lambda_p
\nonumber\\[-8pt]\\[-8pt]
&&\hspace*{17.7pt}{} + N^+_{p,t_{p,n(p)}} \log\lambda_p + N^-_{p,t_{p,n(p)}} \log
\mu_p\bigr] + \mathrm{const},
\nonumber
\end{eqnarray}
where the number of jumps up $N^+_{p,t_{p,n(p)}}:= \sum_{i\ge0}
n^p_{i,i+1}$, the number of jumps down $N^-_{p,t_{p,n(p)}}:= \sum_{i\ge0}
n^p_{i,i-1} $, and the total particle-time
\[
R_{p,t_{p,n(p)}}:=
\int_{t_0}^{t_{p,n(p)}} X_s \,ds = \sum_{i=0}^{\infty} i d^p(i)
\]
for $p=1,\ldots,m$, are the sufficient statistics.\vspace*{1pt} Equation
(\ref{eq2paramFullLike}) shows that, for the E-step, the only
expectations we need are $E_{\tilde{\bg}} [ N^+_{p,t_{p,n(p)}} \mid
\bY]$, $E_{\tilde{\bg}} [ N^-_{p,t_{p,n(p)}} \mid \bY ]$ and
$E_{\tilde{\bg}} [ R_{p,t_{p,n(p)}} \mid \bY]$ for all values
$\tilde{\bg}$. Using independence of the $p$ BDRI processes, the Markov
property and additivity of expectations, we break the desired
expectations into sums of expectations of the numbers of jumps up and
down and the total particle time during each time interval
$[t_{p,k},t_{p,k+1}]$, conditional on $X_{p,t_{p,k}}$ and
$X_{p,t_{p,k+1}}$. By the homogeneity of each of the BDRI processes, in
order to complete the E-step of the EM algorithm, we need to be able to
calculate
%
\begin{eqnarray}
\label{eqcondexpect} U_{i,j}(t) &=& U_{i,j}(t; \lambda, \mu) =
\mathrm{E} \bigl(N^+_t \mid X_0 = i, X_t=j
\bigr),
\nonumber
\\
D_{i,j}(t) &=& D_{i,j}(t; \lambda,\mu) = \mathrm{E}
\bigl(N^-_t \mid X_0 = i, X_t=j \bigr)\quad
\mbox{and}
\\
P_{i,j}(t) &=& P_{i,j}(t;\lambda,\mu) = \mathrm{E}
(R_t \mid X_0 = i, X_t=j )
\nonumber
\end{eqnarray}
for all nonnegative integers $i$ and $j$.

Following \citet{MSlabeledtransitions07}, we choose to work with
restricted moments
%
\begin{eqnarray}
\tilde{U}_{i,j}(t) &=& \tilde U_{i,j}(t; \lambda, \mu) =
\mathrm{E} \bigl(N^+_t 1_{\{X_t=j\}}\mid X_0 = i
\bigr),
\nonumber
\\
\tilde{D}_{i,j}(t) &=& \tilde D_{i,j}(t; \lambda,\mu) = \mathrm
{E} \bigl(N^-_t 1_{\{X_t=j\}} \mid X_0 = i \bigr)\quad
\mbox{and}
\\
\tilde{P}_{i,j}(t) &=& \tilde P_{i,j}(t;\lambda,\mu) = \mathrm
{E} (R_t 1_{\{X_t=j\}} \mid X_0 = i ),
\nonumber
\end{eqnarray}
that we can divide by transition probabilities $p_{i,j}(t)$ to recover the
conditional expectations (\ref{eqcondexpect}),
%
\begin{eqnarray}
\label{eqcond-joint-moments} U_{i,j}(t) &=& \tilde{U}_{i,j}(t)
/ p_{i,j}(t),
\nonumber
\\
D_{i,j}(t) &=& \tilde{D}_{i,j}(t) / p_{i,j}(t) \quad\mbox{and}
\\
P_{i,j}(t) &=& \tilde{P}_{i,j}(t) / p_{i,j}(t).
\nonumber
\end{eqnarray}
In order to compute the restricted moments, we first consider the joint
generating function
%
\begin{equation}
H_i(u,v,w,s,t):= \mathrm{E} \bigl(u^{N^+_t} v^{N^-_t}
e^{-w R_t} s^{X_t} \mid X_0=i \bigr),
\end{equation}
where $0 \le u,v,s \le1$ and $w \ge0$. Partial derivatives of this function,
%
%
\begin{eqnarray}
\label{eqpowerseries} \frac{\partial H_i (u,1,0,s,t )}{\partial u}
\bigg\vert_{u=1} &=& \sum
_{j=0}^\infty s^j \sum
_{n=0}^\infty n \operatorname{Pr}_i
\bigl(N^+_t = n, X_t=j\bigr) \nonumber\\
&=& \sum
_{j=0}^{\infty} \tilde{U}_{i,j}(t)
s^j,
\nonumber
\\
\frac{\partial H_i (1,v,0,s,t )}{\partial v} \bigg\vert_{v=1} &=& \sum
_{j=0}^\infty
s^j \sum_{n=0}^\infty n
\operatorname{Pr}_i\bigl(N^-_t = n, X_t=j
\bigr) \nonumber\\
&=& \sum_{j=0}^{\infty}
\tilde{D}_{i,j}(t) s^j \quad\mbox{and}
\\
\frac{\partial H_i (1,1,w,s,t )}{\partial w} \bigg\vert_{w=0} &=& - \sum
_{j=0}^\infty
s^j \int_0^\infty x \,d
\operatorname{Pr}_i(R_t \le x, X_t=j) \nonumber\\
&=& -
\sum_{j=0}^{\infty} \tilde{P}_{i,j}(t)
s^j
\nonumber
\end{eqnarray}
are power series with coefficients $\tilde{U}_{i,j}(t)$,
$\tilde{D}_{i,j}(t)$ and $-\tilde{P}_{i,j}(t)$, respectively, for
$j=0, 1,\ldots, \infty$, where $\operatorname{Pr}_i$ denotes probability
conditional on
$X_0=i$. We will denote these power series by
$G^+_i(t,s)$, $G^-_i(t,s)$ and $G^*_i(t,s)$,
respectively. If we can compute $G^+_i(t,s)$, $G^-_i(t,s)$ and
$G^*_i(t,s)$ for every possible $t$ and $s$, then we should be able to
recover coefficients of the corresponding power series via differentiation
or integration. Numerical evaluation of the partial derivatives
(\ref{eqpowerseries}) is straightforward if we can compute finite
differences of $H_i(u,v,w,s,t)$. Remarkably, $H_i(u,v,w,s,t)$ is available
in closed form, as we demonstrate in the theorem below, so one can even
obtain derivatives (\ref{eqpowerseries}) analytically. Note that the
theorem below
applies to a general linear BDI process, not only to the BDRI processes.

\begin{theorem}
\label{thmEM}
Let $\{ X_t \}$ be a linear BDI process with parameters $\la\ge0$,
$\mu\ge0$ and
$\nu\ge0$. Over the interval $ [ 0,t ]$, let $N^+_t$ be
the number of jumps
up, $N^-_t$ be the number of jumps down and $R_t$ be the total
particle-time. Then
$H_i(u,v,w,s,t) = E (u^{N^+_t} v^{N^-_t} e^{-w R_t} s^{X_t}\mid X_0=i
)$ satisfies the
following partial differential equation:
%
\begin{equation}
\label{eqPDE} \frac{\partial}{\partial t} H_i = \bigl[ s^2 u
\lambda- (\lambda+ \mu+ w)s + v \mu\bigr] \,\frac{\partial}{\partial s}
H_i +
\nu(us-1) H_i,
\end{equation}
subject to initial condition $H_i(u,v,w,s,0) = s^i$. The Cauchy problem
defined by equation (\ref{eqPDE}) and the initial condition has a unique
solution. When $\lambda> 0$, the solution is
%
\begin{eqnarray}
\label{eqPDEsoln} H_i(u,v,w,s,t) &=& \biggl( \frac{\aa- \alpha_2
({s-\aa
})e^{- \la(\alpha_2 -\aa)ut}/({s-\alpha_2})} {
1- ({s-\aa})
e^{- \la(\alpha_2 -\aa)ut}/({s-a2})}
\biggr)^i
\nonumber\\[-8pt]\\[-8pt]
&&{} \times\biggl( \frac{\aa- \alpha_2}{s- \alpha_2 - (
s-\aa)
e^{- \la(\alpha_2 -\aa)ut}} \biggr)^{{\nu}/{\lambda}} e^{- \nu(1-u
\aa)t},
\nonumber
\end{eqnarray}
where\vspace*{1pt} $\alpha_1 = \frac{\lambda+ \mu+w - \sqrt{ (\lambda+ \mu+
w)^2 - 4\lambda\mu
u v}}{2 \lambda u}$
and
$\alpha_2 = \frac{\lambda+ \mu+w + \sqrt{ (\lambda+ \mu+ w)^2 -
4\lambda\mu
u v}}{2 \lambda u}$.\break  When $\lambda=0$, the solution is
%
\begin{eqnarray}\quad
H_i(u,v,w,s,t) &=& \biggl(s e^{-(\mu+w)t}- \frac{v \mu(e^{-(\mu+w)t} -1
)}{\mu+ w}
\biggr)^i
\nonumber\\[-8pt]\\[-8pt]
&&{} \times e^{{\nu u[v\mu-(\mu+w)s] (e^{-(\mu+w)t}
-1 )}/{(\mu+w)^2} +
\nu({uv\mu}/({\mu+w})-1 )t}.
\nonumber
\end{eqnarray}
\end{theorem}
\begin{pf}
Our proof, detailed in Appendix A,
is a generalization of Kendall's derivation of the
generating function of $X_t$ [\citet{DossSupplement2013,Kendall1948}].
\end{pf}

Having $H_i$ in closed form gives us access to functions $G_i^+$, $G_i^-$
and $G_i^*$, so we are left with the task of recovering coefficients of these
power series. One way to accomplish this task is to differentiate the power
series repeatedly, for example, $\tilde{U}_{i,j}(t) = \frac{1}{j!}\,
\frac{\partial^j
G_i^+(s,t)}{\partial s^j} \vert_{s=0}$. In Appendix C, we demonstrate
that for the death-immigration model ($\lambda= 0$, $\nu\ne0$, $\mu
\ne
0$) and the 
BDRI model considered by \citet{Holmes2005EMindel},
these derivatives can be found analytically [\citet
{DossSupplement2013}]. In general,
repeated differentiation of $G_i^+$, $G_i^-$ and $G_i^*$
needs to be done numerically,
making this method impractical. Instead,
we extend $G_i^+ (t, \cdot)$, $G_i^- (t, \cdot
)$ and
$G_i^* (t, \cdot)$
to the boundary of a unit circle in the complex plane by the change of variables
$s = e^{2 \pi i z}$ ($i$ in this context is the imaginary number $\sqrt
{-1}$, not the initial state of the
BDI process). For example,
\[
G_l^+\bigl(t,e^{2\pi i z}\bigr) = \sum
_{j=0}^{\infty} \tilde{U}_{l,j}(t)e^{2
\pi i j z}
\]
is a periodic function in $z$, which means that $\tilde{U}_{l,j}(t)$ are
Fourier coefficients of this periodic function. Therefore, we can use the
Riemann approximation to the Fourier transform integral to obtain
\[
\tilde{U}_{l,j}(t) = \int_0^1
G_l^+ \bigl(t, e^{2 \pi i s} \bigr)e^{-2 \pi
i j s} \,ds \approx
\frac{1}{K} \sum_{k=0}^{K-1}
G_l^+ \bigl(t, e^{2 \pi i k/K} \bigr)e^{-2
\pi i j k/K}
\]
for some suitably large $K$. The Fast Fourier
Transform (FFT) [\citet{Henrici1979}] can be applied to quickly compute
multiple Fourier coefficients [\citet
{Lange1982fft,Dorman2004,Suchard2008}]. We
do not, however, use the FFT in our algorithm because, for a particular time
interval length $t$, we almost always need to compute $\tilde{U}_{i,j}(t)$,
$\tilde{D}_{i,j}(t)$, $\tilde{P}_{i,j}(t)$ for only one value of $j$.

Now, we can put the pieces together to compute $\mathrm{E}_{\tilde\bg}
[ l_c(\mathbf{X}; \bg) \mid \bY]$. As mentioned above,
$N_{p,t_{p,n(p)}}^+$
equals the sum of the number of jumps up over the disjoint intervals
$[t_{p,i-1}, t_{p,i})$, $i=1,\ldots,n(p)$. The Markov property says that
the conditional expectations of the number of jumps up of $X_{p,t}$ over
$[t_{p,i-1}, t_{p,i})$ given $\bY$ is equal to the conditional expectation
of the number of jumps up over $[t_{p,i-1}, t_{p,i})$ given just
$X_{p,t_{p,i-1}}$ and $X_{p,t_{p,i}}$. Using\vspace*{1pt} similar logic for $N^-_{p,
t_{p,n(p)}}$ and $R_{p,t_{p,n(p)}}$, this gives for $p=1,\ldots,m$,
%
\begin{eqnarray}
\label{eqE-sum-markovprop} \mathrm{E}_{\tilde{\bolds{\gamma}}_p} \bigl[
N^+_{p, t_{p, n(p)}} | \bY\bigr] 
& = &\sum
_{i=1}^{n(p)} U_{X_{p,t_{p,i-1}}, X_{p,t_{p,i}}} (t_{p,i}-t_{p,i-1};
\tilde{\lambda}_p, \tilde{\mu}_p ),
\nonumber
\\
\mathrm{E}_{\tilde{\bolds{\gamma}}_p} \bigl[ N^-_{p, t_{p, n(p)}} |
\bY\bigr] & = & \sum
_{i=1}^{n(p)} D_{X_{p,t_{p,i-1}}, X_{p,t_{p,i}}}
(t_{p,i}-t_{p,i-1}; \tilde{\lambda}_p, \tilde{
\mu}_p ) \quad\mbox{and}
\\
\mathrm{E}_{\tilde{\bolds{\gamma}}_p} [ R_{p, t_{p, n(p)}} | \bY] & = &
\sum
_{i=1}^{n(p)} P_{X_{p,t_{p,i-1}}, X_{p,t_{p,i}}} (t_{p,i}-t_{p,i-1};
\tilde{\lambda}_p, \tilde{\mu}_p ),
\nonumber
\end{eqnarray}
where $\log\tilde\lambda_p = \mathbf{z}_{p,\la}' \tilde\bg
_{p,\lambda}$
and $\log\tilde\mu_p = \mathbf{z}_{p,\mu}' \tilde\bg_{p,\mu}$.
Thus, by\vspace*{1pt}
(\ref{eq2paramFullLike}), (\ref{eqcond-joint-moments}) and
(\ref{eqE-sum-markovprop}), we see that, up to an additive constant, $
\mathrm{E}_{\tilde\bg} [ l_c(\mathbf{X}; \bg) | \bY
]$ is equal
to
\begin{eqnarray*}
&&\sum_{p=1}^m \Biggl\{-t_{n(p)}
\beta\lambda_p \\
&&\qquad\hspace*{0pt}{} + \sum_{i=1}^{n(p)}
\biggl( - \frac{\tilde{P}_{X_{p,t_{p,i-1}}, X_{p,t_{p,i}}}
(t_{p,i}-t_{p,i-1}; \tilde{\lambda}_p, \tilde{\mu}_p
)}{p_{X_{p,t_{p,i-1}}, X_{p,t_{p,i}}} (t_{p,i}-t_{p,i-1}; \tilde
{\lambda}_p, \tilde{\mu}_p )} (\lambda_p + \mu_p)
\\
&&\qquad\hspace*{34.3pt}{}+ \frac{\tilde{U}_{X_{p,t_{p,i-1}}, X_{p,t_{p,i}}}
(t_{p,i}-t_{p,i-1}; \tilde{\lambda}_p, \tilde{\mu}_p
)}{p_{X_{p,t_{p,i-1}}, X_{p,t_{p,i}}} (t_{p,i}-t_{p,i-1}; \tilde
{\lambda}_p, \tilde{\mu}_p )} \log\lambda_p \\
&&\qquad\hspace*{46.3pt}{}+ \frac{\tilde
{D}_{X_{p,t_{p,i-1}}, X_{p,t_{p,i}}}
(t_{p,i}-t_{p,i-1}; \tilde{\lambda}_p, \tilde{\mu}_p
)}{p_{X_{p,t_{p,i-1}}, X_{p,t_{p,i}}} (t_{p,i}-t_{p,i-1}; \tilde
{\lambda}_p, \tilde{\mu}_p )} \log
\mu_p \biggr) \Biggr\},
\end{eqnarray*}
where\vspace*{1pt} the transition probabilities $p_{X_{p,t_{p,i-1}}, X_{p,t_{p,i}}}
(t_{p,i}-t_{p,i-1}; \tilde{\lambda}_p, \tilde{\mu}_p
)$ can be
calculated by using the (known) generating function for the BDI
process, as
is described in Appendix A [\citet{DossSupplement2013}].

\subsection{M-step}
To complete the M-step 
for each iteration of the
EM algorithm, we use a Newton--Raphson algorithm to maximize
\[
f(\bg) = \mathrm{E}_{\tilde{\bg}} \bigl[ l_c(\mathbf{X}; \bg) |
\bY\bigr].
\]
In
each
Newton--Raphson step, we update $\bg$ via the following recursion:
\[
\bg_{\mathrm{new}} = \bg_{\mathrm{cur}} - \bigl[ \mathbf{H} f(\bg
_{\mathrm{cur}}) \bigr]^{-1} \bolds{\nabla} f(
\bg_{\mathrm{cur}}),
\]
where $\bolds{\nabla} f(\bg_{\mathrm{cur}})$ is the gradient
vector and
$\mathbf{H} f(\bg_{\mathrm{cur}})$ is the Hessian matrix of the function
$f(\bg)$. If we collect the observation times into a vector $\mathbf
{T}' =
(t_{1,n(1)},\ldots, t_{m,n(m)})$, the expectations of the sufficient
statistics into vectors
%
\begin{eqnarray}
\label{eqdefnUDP} \mathbf{U}' &=& \bigl(E_{\tilde{\bg}} \bigl[
N^+_{1,t_{1,n(1)}} | \bY\bigr],\ldots, E_{\tilde{\bg}} \bigl[
N^+_{m,t_{m,n(m)}} | \bY\bigr] \bigr),
\nonumber
\\
\mathbf{D}' &=& \bigl(E_{\tilde{\bg}} \bigl[ N^-_{1,t_{1,n(1)}} |
\bY\bigr],\ldots, E_{\tilde{\bg}} \bigl[ N^-_{m,t_{m,n(m)}} | \bY\bigr
] \bigr),
\\
\mathbf{P}' &=& \bigl(E_{\tilde{\bg}} [ R_{1,t_{1,n(1)}} | \bY],\ldots, E_{\tilde{\bg}} [ R_{m,t_{m,n(m)}} | \bY] \bigr),
\nonumber
\end{eqnarray}
and the process-specific birth and death rates into vectors
\[
\bolds{\lambda}' = (\lambda_1,\ldots,
\lambda_m) \quad\mbox{and}\quad \bolds{\mu}' = (
\mu_1,\ldots, \mu_m),
\]
then after defining covariate matrices
\[
\mathbf{Z}_{\lambda}' = (\mathbf{z}_{1,\lambda},\ldots,
\mathbf{z}_{m,\lambda}) \quad\mbox{and}\quad \mathbf{Z}_{\mu}'
= (\mathbf{z}_{1,\mu},\ldots,\mathbf{z}_{m,\mu}),
\]
the gradient and the Hessian can be compactly expressed in matrix form as
%
\begin{eqnarray}\qquad
\bolds{\nabla} f(\bg) & = & \bigl( \mathbf{Z}_{\lambda}'
\bigl[-\operatorname{diag}(\mathbf{P} + \beta\mathbf{T})\bolds
{\lambda} +
\mathbf{U} \bigr], \mathbf{Z}_{\mu}' \bigl[-
\operatorname{diag}(\mathbf{P})\bolds{\mu} + \mathbf{D} \bigr]
\bigr),
\\
\mathbf{H} f(\bg) &=& \pmatrix{ - \mathbf{Z}_{\lambda}'
\operatorname{diag}(\mathbf{P} + \beta\mathbf{T})\operatorname
{diag}(\bolds{
\lambda})\mathbf{Z}_{\lambda} & \mathbf{0}
\vspace*{2pt}\cr
\mathbf{0} & -
\mathbf{Z}_{\mu}' \operatorname{diag}(\mathbf{P})
\operatorname{diag}(\bolds{\mu})\mathbf{Z}_{\mu}},
\end{eqnarray}
which we show in Appendix B; see (S-4), (S-6) and (S-9) [\citet
{DossSupplement2013}]. Notice
that the algebraic separation of the birth and the death components in the
complete-data likelihood results in\vspace*{1pt} blocks---corresponding to
$\bg_{\lambda}$ and $\bg_{\mu}$---in the above formulae. The fact
that the
gradient and Hessian of $f(\bg)$ is available analytically results in fast
execution of Newton--Raphson updates. In our experience, the Newton--Raphson
algorithm in our M-step converges after only 3--5 iterations. However,
we also
note that it is not critical to achieve convergence of this algorithm since
even a single Newton--Raphson update within the M-step is enough to guarantee
the usual convergence properties of the EM algorithm
[\citet{Lange1995gradientEM}].

We obtain the observed Fisher information via Louis' formula:
\[
\hat{I}_{\mathbf{Y}}(\hat{\bg}) = \mathrm{E}_{\hat{\bg}} \bigl[ -
\mathbf{H} l_c(\mathbf{X};\hat{\bg}) | \mathbf{Y} \bigr] -
\mathrm{E}_{\hat
{\bg}} \bigl[ \bolds{\nabla} l_c(\mathbf{X};
\hat{\bg}) \bolds{\nabla} l_c(\mathbf{X};\hat{
\bg})' | \mathbf{Y} \bigr],
\]
where
$\bolds{\nabla} l_c$ is the gradient and $\mathbf{H} l_c$ is
the Hessian
of the complete-data log-likelihood [\citet{Louis1982}]. This requires
calculation of the conditional cross-product means, $\mathrm{E} [ N^+_t
N^-_t \mid\bY]$, $\mathrm{E} [ N^+_t R_t | \bY]$,
$\mathrm{E} [ N^-_t R_t | \bY]$, and the conditional
second moments
of $N^+_t,N^-_T$ and $R_t$. The derivation of the information in terms of
these moments is in Appendix B [\citet{DossSupplement2013}]. These
conditional second- and cross-moments,
as well as $\mathbf{P}$ and $\mathbf{D}$, can be computed in
analogous fashion to
$\mathbf{U}$ above, using the joint generating function (\ref{eqPDEsoln}).
We use the information matrix to compute approximate standard errors of
$\hat{\bg}$
and use these standard errors together with asymptotic normality of
maximum likelihood estimators
to form confidence intervals and sets for our model parameters.

\section{Results}

\subsection{Simulations}
To test our methods, we simulate data from the BDRI model with $\lambda=0.07$,
$\mu=0.12$ and $\beta= 1.2$, where $\beta$ is assumed to be known,
leaving us
with only two parameters to estimate: $\lambda$ and $\mu$. We choose these
parameters to resemble, but not exactly match, the dynamics of our biological
example, discussed in the next subsection. We simulate
$100$
independent processes starting from initial states drawn uniformly between
$1$ and $15$. From each process we collect at least two observations. We
place observation times uniformly between $0$ and $30$. Table~\ref
{tabsummStats} gives some summary statistics for the simulated data.

%
\begin{table}
\tablewidth=280pt
\caption{Summary statistics for the simulated and \textit{M.
tuberculosis} IS\textit{6110} data}
\label{tabsummStats}
\begin{tabular*}{\tablewidth}{@{\extracolsep{\fill}}ld{4.1}d{3.2}@{}}
\hline
\textbf{Value} &
\multicolumn{1}{c}{\textbf{Simulated data}} &
\multicolumn{1}{c@{}}{\textbf{IS\textit{6110} data}} \\
\hline
Number of intervals &
387 &
252 \\
Average interval length &
5 &
0.35 \\
Number of individuals &
100 &
196 \\
Number of intervals with an increase &
78 &
14 \\
Average increase given an increase &
1.5 &
1 \\
Number of intervals with a decrease &
190 &
14 \\
Average decrease given a decrease &
2.5 &
1.2 \\
Number of intervals with no change &
119 &
224 \\
Mean starting state &
5.5 &
11 \\
Standard deviation of starting state &
3.8 &
5.3 \\
Total length of time &
1947 &
89 \\
\hline
\end{tabular*}
\end{table}

We test our EM algorithm and confidence interval calculations on these
simulated data with initial parameter values of $0.2$ for both $\lambda
$ and
$\mu$. We considered other choices of starting values, but the
algorithm was
not sensitive to them. Notice that this is the simplest
parameterization of
our BDRI model, where both $\mathbf{z}_{\la}$ and $\mathbf{z}_{\mu
}$ are
vectors of ones. We estimate $0.067$ with a 95\%
confidence interval of $ (
0.052,0.081
)$ for $\lambda$ and $0.12, (
0.1,0.14
)$ for $\mu$, indicating that our algorithm successfully recovered
these BDRI model parameters. We also conduct a similar simulation
study for the BDRI model with covariates, successfully estimating
parameters of
this model as well, but omit detailed results of this simulation for brevity.
%
%

\subsection{Comparison with the frequent monitoring method}
\begin{figure}

\includegraphics{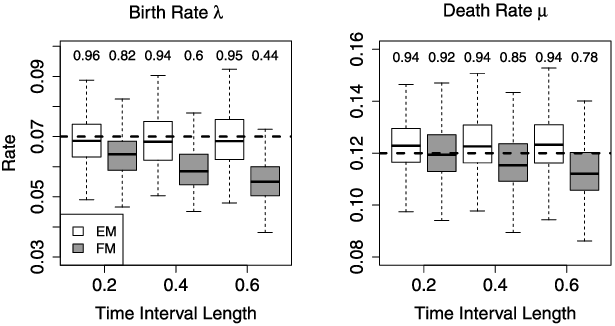}

\caption{Box plots of birth (left panel) and
death (right panel) rate estimates, obtained from 200 simulated data sets
using the EM algorithm and frequent monitoring (FM) method. The true
parameter values, used in
data simulations, are marked by the horizontal dashed lines. Above the
box plots, we show Monte Carlo estimates
of coverage probabilities of the 95\% confidence intervals.}
\label{figcomparisonPlotsL}\vspace*{-3pt}
\end{figure}

We compare our EM algorithm for computing the actual MLE to the
frequent monitoring (FM) method of \citet{RosenbergTrans03} for
computing the MLE of an approximate likelihood. In the FM
method, \citet{RosenbergTrans03} assume that if the
starting and ending values of the birth--death process are equal for a
particular interval,
then no jumps occurred in this interval. Further, if the difference
between the starting and ending values is $-1$ or $1$, then
exactly one jump up or exactly one jump down must have occurred, respectively.
The authors exclude all observed intervals, for which starting and
ending values
differ by more than one unit. Let $i$ be
the starting state for an interval, $t$ the length of the interval
and $\lambda_i = i (\lambda+ \mu)$. Then the corresponding
probabilities for the three possible events are $e^{-\lambda_i u}$,
$\frac{i \lambda}{\lambda_i} (1 - e^{-\lambda_i u} )$
and $
\frac{i\mu}{\lambda_i} (1-e^{-\lambda_i u} )$, respectively.\vspace*{2pt}
%
\citet{RosenbergTrans03} use this FM method to estimate rates in what is
effectively a multi-state branching process, but we will compare the two
methods on our BDRI model with the immigration rate $\beta$
constrained to be
$0$. We again simulate an underlying BD process using $\lambda=
0.07$ and $\mu= 0.12$.
To compare the two methods, we generate three different sets of data. In
each set, we generate observed states of the BD process at a fixed constant
distance $dt$ apart. This distance varies across the data sets, taking the
values $0.2,0.4$ and $0.6$, respectively. We repeat this procedure 200 times
and compute birth and death rate estimates and corresponding 95\% confidence
intervals using the EM algorithm and FM approximation method. We show box
plots of the resulting estimates for $\lambda$ and $\mu$ in
Figure \ref{figcomparisonPlotsL}. As expected, the FM estimates behave
reasonably when interval lengths are small, but the approximation becomes
poor as we increase the interval length. The FM method always underestimates
the parameters since the method effectively undercounts the number of
unobserved jumps in the BD process. We also compute Monte Carlo
estimates of
coverage probabilities of the two methods, shown above the box plots in
Figure \ref{figcomparisonPlotsL}. Not surprisingly, coverage of the
95\%
confidence intervals computed under the proper BD model likelihood are very
close to the promised value of 0.95. In contrast, the FM approximation-based
95\% confidence intervals contain the true parameter value less than
95\% for
all three simulation scenarios.

\subsection{Mycobacterium tuberculosis IS\textit{6110} transposon}
We apply our EM algorithm to estimation of birth and death rates of the
transposon IS\textit{6110} in \textit{M. tuberculosis} [\citet
{McEvoy2007}]. A
transposon, or transposable element, is a genetic sequence that can
duplicate, remove itself and jump to a new location in the genome.
IS\textit{6110} is a transposon that plays an important role in
epidemiological studies of tuberculosis. More specifically, the number and
locations of IS\textit{6110} elements in the \textit{M. tuberculosis}
form a
genetic signature or genotype of the mycobacterium, allowing epidemiologists
to draw inference about disease transmission when the same genotype is
observed among patients with active tuberculosis [\citet
{vanEmbden1993}]. Such
genotypic comparison can translate into meaningful epidemiological inference
only if the dynamics of IS\textit{6110} evolution are well understood.
Therefore, accurate estimation of rates of changes of IS\textit{6110}-based
genotypes is critical for using these genotypes in epidemiological studies
[\citet{Tanaka2001}].

We analyze data from an ongoing population-based study that includes all
tuberculosis cases reported to the San Francisco Department of Public Health
[\citet{CattamanchiEtAl2006Tuberculosis}]. Our data include patients
with more
than one \textit{M. tuberculosis} isolate from specimens sampled more
than 10
days apart and genotyped with IS\textit{6110} restriction fragment length
polymorphism. We ignore genomic locations of IS\textit{6110} and
assume that
the transposon counts are discretely observed realizations of a BDRI process,
with no immigration ($\beta= 0$); in particular, we assume that
patients are not
reinfected with a different strain of the bacteria in the period between
observations. The third column in Table \ref{tabsummStats} gives
summary statistics for the
data.

We first use a simple model with one single birth rate and one single death
rate of the IS\textit{6110} for all patients.
In the analysis presented, we start the EM algorithm with parameter
guesses of $0.05$ and $0.05$ for $\lambda$ and $\mu$, respectively, and
their\vadjust{\goodbreak}
MLEs are
$0.0176$ and $0.0207$, respectively. The starting values for the EM do
not affect these results. Our estimate and 95\% confidence interval for
$\lambda$, $0.0176$ and $(0.0082, 0.027)$, are consistent with the
corresponding quantities, $0.0188$ and $(0.0085, 0.0291)$, from
\citet{RosenbergTrans03}. Although the authors' confidence
interval for $\mu$, $(0.0057, 0.0237)$, overlaps with ours, $(0.011,
0.031)$, our estimate for $\mu$, $0.0207$, is noticeably higher than
Rosenberg, Tsolaki and Tanaka's (\citeyear{RosenbergTrans03}) estimate
of $0.0147$. Note from Table \ref {tabsummStats} that among the
intervals with a decrease, the average count drop is by more than $1$;
there are $3$ intervals where IS\textit{6110} counts drop by $2$,
whereas there are no intervals that experience an increase by more than
$1$. Thus, we would expect our estimate for $\mu $ to increase over
Rosenberg, Tsolaki and Tanaka's (\citeyear{RosenbergTrans03}) approximation, whereas that of
$\lambda$ should be similar between the two methods. We also point out
that we analyze an updated version of the data analyzed by
\citet{RosenbergTrans03}. Moreover, \citet{RosenbergTrans03}
use a slightly more complicated model for IS\textit{6110} evolution,
which takes into account shifts in transposon location. We conclude
that estimates of birth and death rates of IS\textit{6110} do not vary
dramatically when estimation methods and data collection are altered.
We now turn to more complicated BDRI models that have not been applied
before to the \textit{M. tuberculosis} IS\textit{6110} copy number
evolution. These models will take into account potential dependence of
IS\textit{6110} birth and death rates on patient-specific covariates.

\subsubsection{\textit{Mycobacterium tuberculosis} lineage comparison}
\begin{figure}

\includegraphics{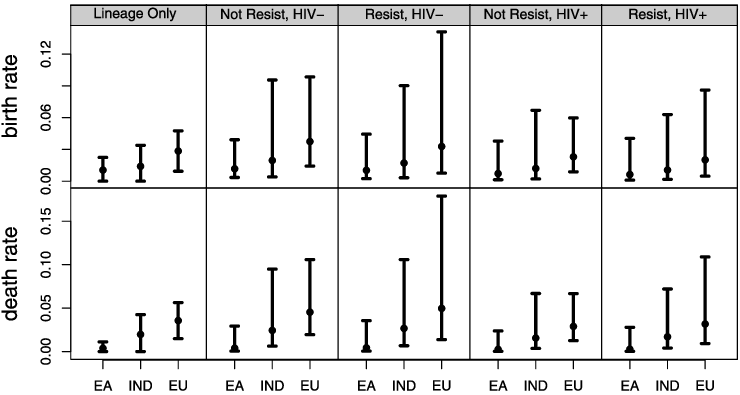}

\caption{Point estimates and 95\% confidence intervals for birth and death
rates of the IS\textit{6110} transposable element obtained by separately
analyzing three \textit{M. tuberculosis} lineages: European--American
(EU), Indo-Oceanic (IND) and East Asian (EA) (leftmost column) and by
fitting the log-linear model with lineage, drug resistance and HIV
status as covariates. For the latter model, the estimated regression
coefficients are transformed into four sets of lineage-specific birth and
death rates (last four columns).}
\label{figcovariateplots1}
\end{figure}

In addition to estimation of the global birth and death rates, we separately
estimate these parameters in each of the three lineages of
\textit{M. tuberculosis} observed in San Francisco. Based on genomic
sequence similarity, \textit{M. tuberculosis} is divided into six main
lineages: Euro-American, East-Asian, Indo-Oceanic, East-African--Indian,
West-African I and West-African II [\citet{Gagneux2006}]. In our
lineage-specific analysis, we consider 109 individuals infected with
Euro-American (EU) lineage strains, 54~individuals infected with East-Asian
(EA) lineage strains and 25 individuals infected with Indo-Oceanic (IND)
lineage strains. One simple way to accommodate this lineage effect is to
build a log-linear model for birth and death rates with two categorical
covariates:
\[
\log\lambda_p = \gamma_{\la,1}
+ \gamma_{\la,2} \operatorname{EU}_p + \gamma_{\la,3}
\operatorname{IND}_p, \log\mu_p = \gamma_{\mu,1} +
\gamma_{\mu,2} \operatorname{EU}_p + \gamma_{\mu,3}
\operatorname{IND}_p,
\]
where $\mathrm{EU}_p=1$ if patient $p$ is infected with the EU strain
and 0
otherwise, and $\mathrm{IND}_p=1$ if patient $p$ is infected with the
IND strain
and 0 otherwise. The intercepts, $\gamma_{\la,1}$ and $\gamma_{\mu,1}$,
correspond to birth and death of the EA strain. We transform the coefficients
$(\gamma_{\la,1}, \gamma_{\la,2}, \gamma_{\la,3})$ and $(\gamma
_{\mu,1},
\gamma_{\mu,2}, \gamma_{\mu,3})$ into the \textit{M. tuberculosis}
lineage-specific birth and death rates and show these estimates
together with
their corresponding confidence in the first column of
Figure~\ref{figcovariateplots1}. Most notably,
%
there appears to be a substantial difference between death rates of the
Euro-American and East-Asian\vadjust{\goodbreak} lineages. We report regression
coefficients on
the multiplicative scale [e.g., $\exp(\gamma_{\la,1}$)] with their
corresponding
95\% confidence intervals in the lineage model columns of
Table \ref{lineagetable}. In this table the highlighted EU rate
multiplier shows that the death rate of
IS\textit{6110} copy number
is estimated to be approximately ten times higher than the
corresponding death rate in the EA lineage.
The confidence interval of the EU rate multiplier does not contain one,
indicating that EA and EU lineages have different death
rates of the IS\textit{6110} transposon.

%
\begin{table}
\caption{Results of the two log-linear models for birth and death
rates of IS\textit{6110}. The lineage model includes only effects
of \textit{M. tuberculosis} lineages (EA, EU, IND). The full model
combines the effects of lineages, HIV infection status ($\mathrm
{HIV}^+$) and drug resistance status (DR). The birth and death rate
multiplier estimates for the EU lineage are highlighted in bold to
indicate that the confidence intervals for these parameters are above one}
\label{lineagetable}
\begin{tabular*}{\tablewidth}{@{\extracolsep{\fill}}l@{\ }llcd{2.3}c@{}}
\hline
& & \multicolumn{2}{c}{\textbf{Lineage model}} & \multicolumn{2}{c@{}}{\textbf{Full
model}}\\[-4pt]
& & \multicolumn{2}{c}{\hrulefill} & \multicolumn{2}{c@{}}{\hrulefill}
\\
\multicolumn{2}{@{}l}{\textbf{Coefficient}} & \multicolumn{1}{c}{\textbf{MLE}}
& \multicolumn{1}{c}{\textbf{CIs}} & \multicolumn{1}{c}{\textbf{MLE}} &
\multicolumn{1}{c@{}}{\textbf{CIs}}\\
\hline EA birth rate, & $\exp(\gamma_{\lambda,1})$ & $0.011$ & (0.003,
0.034) & 0.012 & (0.006, 0.025)
\\
EU multiplier, & $\exp(\gamma_{\lambda,2})$ & $2.63$ & (0.689, 10.0)
& \multicolumn{1}{c}{\textbf{3.2}\hphantom{0}} & \textbf{(1.1, 9.4)}
\\
IND multiplier, & $\exp(\gamma_{\lambda,3})$ & $1.40$ & (0.229, 8.53)
& 1.7 & (0.29, 9.7)
\\
DR multiplier, & $\exp(\gamma_{\lambda,4})$ & \multicolumn
{1}{c}{--} & \multicolumn{1}{c}{--} & 0.88 & (0.36, 2.1)
\\
$\mathrm{HIV}^+$ multiplier, & $\exp(\gamma_{\lambda,5})$ &
\multicolumn{1}{c}{--} & \multicolumn{1}{c}{--} & 0.61 & (0.28, 1.3)
\\
EA death rate, & $\exp(\gamma_{\mu,1})$ & $0.004$ & (0.0005, 0.028)
& 0.004 & (0.0005, 0.031)
\\
EU multiplier, & $\exp(\gamma_{\mu,2})$ & \textbf{9.32} &
\textbf{(1.19, 72.8)} & \multicolumn{1}{c}{\textbf{11}\hphantom{00\textbf{..}}} & \textbf{(1.2, 114)}
\\
IND multiplier, & $\exp(\gamma_{\mu,3})$ & $5.40$ & (0.553, 52.6) &
6.2 & (0.36, 1.1)\\
DR multiplier, & $\exp(\gamma_{\mu,4})$ & \multicolumn{1}{c}{--} &
\multicolumn{1}{c}{--} & 1.1 & (0.52, 2.3)
\\
$\mathrm{HIV}^+$ multiplier, & $\exp(\gamma_{\mu,5})$ &
\multicolumn{1}{c}{--} & \multicolumn{1}{c}{--} & 0.64 & (0.36, 1.1)
\\
\hline
\end{tabular*}
\end{table}

Since this is a novel result that has implications for monitoring
tuberculosis with molecular genotyping,
we examine the difference in death rates between the three lineages
more closely. More specifically, we add two binary
covariates to our log-linear model: \textit{M. tuberculosis} drug
resistance (DR) and HIV infection status of each
patient ($\mathrm{HIV}^+$). Our new model for birth and death rates becomes
\begin{eqnarray*}
\log\lambda_p &=& \gamma_{\la,1} +
\gamma_{\la,2} \operatorname{EU}_p + \gamma_{\la,3}
\operatorname{IND}_p + \gamma_{\la,5} \operatorname{DR}_p
+ \gamma_{\la,4} \operatorname{HIV}^+_p,
\\
\log\mu_p &=& \gamma_{\mu,1} + \gamma_{\mu,2}
\operatorname{EU}_p + \gamma_{\mu,3} \operatorname{IND}_p
+ \gamma_{\mu,5} \operatorname{DR}_p + \gamma_{\mu,4}
\operatorname{HIV}^+_p,
\end{eqnarray*}
where $\mathrm{DR}_p=1$ if patient $p$ is infected with a drug
resistant strain
\textit{M. tuberculosis} and 0 otherwise, and $\mathrm{HIV}^+_p=1$ if patient
$p$ is infected with HIV and 0 otherwise. Parameter estimates of this full
model and their corresponding 95\% confidence intervals are reported in the
full model columns of Table~\ref{lineagetable}.
The HIV infection and\vadjust{\goodbreak}
drug resistance appear to have no effect on the birth and death rates of
IS\textit{6110} transposon.
IS\textit{6110} copy number variation may have an impact on functions
of neighboring genes
in the \textit{M. tuberculosis} genome [\citet{Alonso2011}].
Therefore, IS\textit{6110} copy number can potentially interact with other
\textit{M. tuberculosis} phenotypes, such as drug resistance and
adaptation to HIV and antiviral treatment,
with the help of selection [\citet{McEvoy2007}]. However, we do not
expect to see association between IS\textit{6110}
copy number and \textit{M. tuberculosis} phenotypes within one patient
because selection is unlikely to
play a role on such a short time scale. Hence, we view our estimated
small effects of HIV infection and drug resistance on
IS\textit{6110} copy number as biologically plausible.
The EU lineage effect on the
death rate remains statistically significant even after controlling for the
two additional covariates. Interestingly, the EU lineage effect on the
birth rate also becomes
statistically significant in the full model.
Effect sizes for both birth and death rates increase and the confidence
intervals include larger values in the full model over the lineage-only
model. This indicates that the full model tends to find more
differences in rates between the
lineages than the lineage-only model does. While more data are certainly
needed to confirm that EU lineage birth rate effect is not $1$, the full
model may be capturing information the simpler lineage-only model does not,
which, in the face of limited data, is valuable. For
practical considerations, the fact that our most parameter rich full model
results in significant effects of EU lineage on IS\textit{6110} birth and
death rates suggests that \textit{M. tuberculosis} lineage has to be
taken into consideration when IS\textit{6110} genotype data
are used to uncover the history of \textit{M. tuberculosis}
transmission.\vadjust{\goodbreak}

\subsubsection{IS\textit{6110} counts}
\begin{figure}

\includegraphics{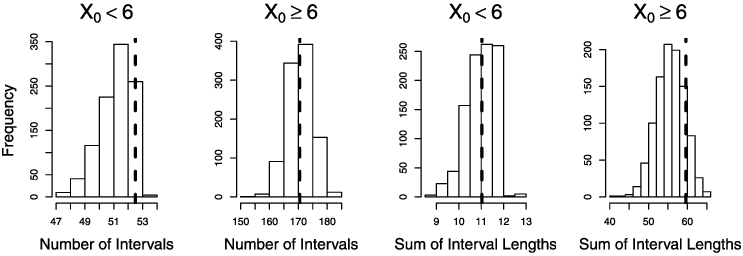}

\caption{Low- vs high-count genotype analysis. Histograms of simulated
numbers of intervals and
sums of interval lengths are plotted for intervals with starting values
less than six and greater
or equal to six. The vertical dashed lines indicate the observed values
of the four statistics.}
\label{figgoodfit}
\end{figure}

The initial number of IS\textit{6110} elements is a potential
confounder in our analysis because patients infected with Euro-American
and East-Asian differ drastically in the number of IS\textit{6110}
elements at the beginning of the observation period. The isolates from
the Euro-American lineage have between 2 and 17 IS\textit{6110}
elements, with 41 out of 109 patients having the first recorded
IS\textit{6110} count less than 6, while IS\textit{6110} counts vary
between 6 and 22 for the East-Asian isolates. \citet{Warren2002}
suggest that IS\textit{6110} genotypes with fewer than six elements
have a very low rate of change, because in their data cases with no
observed changes in the genotype are dominated by such low-count
genotypes. However, our birth--death model very well predicts the
conclusion of \citet{Warren2002} that low-count genotypes evolve
slower than high-count genotypes. To demonstrate this, we simulate 1000
data sets using our global birth and death rates and observed initial
IS\textit{6110} counts for each patient. We record the number of
intervals with equal starting and ending values less than six,
$n_{0,<6}$, and equal starting and ending values greater or equal to
six, $n_{0, \ge6}$. We also recorded the length sum of both kinds of
intervals: $t_{0, <6}$ and $t_{0, \ge6}$. In our data,
$n_{0,<6}^{\mathrm{obs}} = 53$ and $n_{0,\ge6}^{\mathrm{obs}} = 171$
with $n_{0, <6}^{\mathrm{obs}}/t_{0,<6}^{\mathrm{obs}} = 4.6 > 2.8 =
n_{0, \ge 6}^{\mathrm{obs}}/t_{0, \ge6}^{\mathrm{obs}}$, in
agreement\vspace*{2pt} with Warren et~al.'s (\citeyear{Warren2002})
analysis. Histograms of simulated values of the four statistics,
$n_{0,<6}$, $n_{0,\ge6}$, $t_{0,<6}$ and $t_{0, \ge6}$, shown in
Figure~\ref{figgoodfit}, demonstrate that our birth--death model replicates
%
well the observed dynamics of low-count and high-count IS\textit{6110}
genotypes. We conclude that our data do not provide evidence that
evolutionary dynamics of low-count genotypes differ from high-count
genotype dynamics. Therefore, it is unlikely that a high percentage of
low-count genotypes in the Euro-American lineage isolates causes our
estimated discrepancy between death rates of Euro-American and
East-Asian \textit{M. tuberculosis} lineages.



\section{Discussion}
\label{secDiscussion}
In this paper we present a novel EM algorithm for fitting birth--death
processes to panel data. We allow logarithms of
birth and death rates to be linear combinations of individual-level
covariates. Such birth--death models with covariates share analogy
with covariate-dependent CTMC models on finite state spaces---a widely
used class of models in medical statistics [\citet{Kalbfleisch1985}].
To our knowledge,
there is no established and well tested method for fitting birth--death
processes, considered in this paper, to panel data.
We hope that by filling this void with our new EM algorithm,
accompanied by an open-source R package \verb8DOBAD8 (available at
\url{http://cran.r-project.org}), we will stimulate statistical
applications of birth--death processes, at least in the context of panel data.

We illustrate the applicability of birth--death models by analyzing the
evolutionary dynamics of the IS\textit{6110} transposon---an important
genetic marker that serves as a genetic signature of the
\textit{M. tuberculosis} bacterium. By building realistic models for
IS\textit{6110} dynamics, we uncover differences in IS\textit{6110}
birth and
death rates among major lineages of \textit{M. tuberculosis}, while
controlling for other clinical covariates. This novel result is important
because IS\textit{6110} copy number is used as a genetic marker to
create DNA
fingerprints of \textit{M. tuberculosis} using the restriction fragment
length polymorphism technology [\citet{vanEmbden1993,KatoMaeda2011}]. Strains
that have the same IS\textit{6110} counts and in which the IS\textit{6110}
element is located in DNA fragments of similar size are considered
identical. When such identical strains are found in community-based studies,
the strains are clustered and
patients carrying these strains are inferred
to belong to the same
\textit{M. tuberculosis} transmission chain [\citet{KatoMaeda2011}].
However, if some
\textit{M. tuberculosis} lineages evolve at much slower rates than
others, as
we discover in our analysis, then using the same notion of similarity between
IS\textit{6110} counts for these slow-evolving lineages could be highly
misleading. Therefore, we suggest that when using IS\textit{6110} genotypes,
\textit{M. tuberculosis} lineage effect should be included explicitly in
statistical protocols of estimating tuberculosis epidemiological
clusters.

Although in our \textit{M. tuberculosis} fingerprinting example we do
not consider the possibility of immigration, we include immigration in
our methodological developments. More specifically, our EM algorithm
and the accompanying software package allow for immigration to occur at
a rate proportional to the birth rate. We have two reasons for
including this generalization. First, this limited form of immigration
complicates neither our mathematical developments nor computational
tractability of the EM algorithm. Second, incorporating immigration
makes our EM algorithm more transferable to other domains of
application of birth--death processes. For example, our methodological
developments directly apply to modeling the evolution of insertions and
deletions in molecular sequences, where immigration is needed to
prevent molecular sequences contracting to length zero
[\citet{TKF91,Holmes2005EMindel}]. Moreover, as we show in
Appendix C, for this particular application, the E-step of our EM
algorithm is available in closed form, eliminating the need for
numerical integration [\citet {DossSupplement2013}]. Another
example of potential transferability of our EM algorithm is for hidden
death-immigration models for recurrent medical conditions, such as that
considered by \citet{Crespi2005}. Although our EM algorithm does
not apply directly to the application these authors consider, because
the states of the immigration-death process are only partially observed
at discrete time points, our mathematical results remain useful here.
More specifically, one can use our mathematical developments in the
context of continuous-time hidden Markov models
[\citet{Roberts2008}] in order to develop an EM algorithm, akin to
a classical Baum--Welch algorithm [\citet{Baum1970}]. As in the
aforementioned insertion-deletion model, Appendix C demonstrates that
the expectations of complete data sufficient statistics for the
death-immigration model are available in closed form
[\citet{DossSupplement2013}]. We note that because our Theorem
\ref{thmEM} applies to general linear BDI models, we are able to use
this theorem to study properties of a death-immigration model, which is
not a BDRI model---the main focus of this manuscript.

Finally, we would like to point out that the generating functions
derived in
Theorem \ref{thmEM} are useful not only for developing EM algorithms for
birth--death models, but also for probabilistic characterization of
birth--death trajectories in general.
For example, we are not aware of analytic formulae for expectations of
the sufficient statistics that do not involve
the ending state of the process at time $t$: $\mathrm{E}(N_t^+ \mid X_0=i)$,
$\mathrm{E}(N_t^- \mid X_0=i)$ and $\mathrm{E}(R_t^+ \mid X_0=i)$. These
expectations, useful for prediction purposes, arise analytically
from the generating functions in Theorem \ref{thmEM} [e.g., $\mathrm{E}(N_t^+
\mid X_0=i) =\partial H_i (u,1,0,1,t )/\partial u \vert
_{u=1} $].




\section*{Acknowledgment}

We thank Peter Guttorp for stimulating discussions and for pointing us
to the work of \citet{Golinelli2000}.

\begin{supplement}
\stitle{Further mathematical details}
\slink[doi]{10.1214/13-AOAS673SUPP} 
\sdatatype{.pdf}
\sfilename{aoas673\_supp.pdf}
\sdescription{Appendices referenced in Sections~\ref{secMethodology}
and \ref{secDiscussion} are available in the supplementary material
[\citet{DossSupplement2013}].}
\end{supplement}


\printaddresses

\end{document}